# Software-Defined Metasurface Paradigm: Concept, Challenges, Prospects

A. Pitilakis [1,2], A.C. Tasolamprou [1], C. Liaskos [1], F. Liu [3], O. Tsilipakos [1], X. Wang [3], M.S. Mirmoosa [3], K. Kossifos [4], J. Georgiou [4], A. Pitsilides [5], N.V. Kantartzis [1,2], S. Ioannidis [1], E.N. Economou [1], M. Kafesaki [1,6], S.A. Tretyakov [3] and C.M. Soukoulis [1,7]

[1] Foundation for Research and Technology Hellas, 71110, Heraklion, Crete, Greece
[2] Department of Electrical and Computer Engineering, Aristotle University of Thessaloniki, Thessaloniki, Greece
[3] Department of Electronics and Nanoengineering, Aalto University, P.O. Box 15500, Espoo, Finland
[4] Department of Electrical and Computer Engineering, University of Cyprus, 20537, Nicosia, Cyprus
[4] Department of Computer Science, University of Cyprus, 20537, Nicosia, Cyprus
[6] Department of Materials Science and Technology, University of Crete, 71003, Heraklion, Crete, Greece
[7] Ames Laboratory and Department of Physics and Astronomy, Iowa State University, Ames, Iowa 50011, USA
Email: alexpiti@ece.auth.gr; atasolam@iesl.forth.gr

*Abstract* – HyperSurfaces (HSFs) are devices whose electromagnetic (EM) behavior is software-driven, i.e., it can be defined programmatically. The key components of this emerging technology are the metasurfaces, artificial layered materials whose EM properties depend on their internal subwavelength structuring. HSFs merge metasurfaces with a network of miniaturized custom electronic controllers, the nanonetwork, in an integrated scalable hardware platform. The nanonetwork receives external programmatic commands expressing the desired end-functionality and appropriately alters the metasurface configuration thus yielding the respective EM behavior for the HSF. In this work, we will present all the components of the HSF paradigm, as well as highlight the underlying challenges and future prospects.

## I. INTRODUCTION

Metasurfaces, thin planar artificial structures, have enabled the realization of novel components with engineered and even unnatural functionalities for the RF, THz and optical spectrum; applications include EM invisibility, total radiation absorption, filtering and steering of impinging waves, as well as ultra-efficient, miniaturized antennas for sensors and implantable communication devices [1,2]. Despite these impressive capabilities, incorporation of metasurfaces to real-world systems is not straightforward: Firstly, they are typically non-adaptive and non-reusable, restricting their applicability to a single functionality per structure; secondly, for flexibility, accessibility and reusability there is a need for a clear hardware and software stack; thirdly, their design remains an interdisciplinary task for specialized researchers. The mission of the HSF platform is to tackle all the aforementioned issues by: hosting multiple concurrent tunable EM functionalities described in a clean and simple application programming interface (API); integrating reconfigurable metasurfaces with electronics enabled by adaptive switch nanonetworks (allowing for global and local control) in a well-packaged, truly smart metamaterial; deploying simple reusable software modules that expose the desired end-functionality to users without specialized metasurface competences. Apart from the development of a working HSF prototype, it is necessary to highlight the challenges and perspectives of this new technology. To this end, we explore the scaling of these smart materials in order to estimate fundamental limitations, stemming from forthcoming nano-electronic and fabrication technologies. Extending the applicability spectrum, we identify the optimum metasurface types for a given need, selecting prime candidates from the large metamaterials family. Overall, the HSF paradigm comes as a timely step towards the introduction of the *Internet of Materials* concept, underpinning the global adaptation of IoT (Internet of Things) systems. The structure of this paper is as follows: Section II presents the HSF architecture and details its three layers (metasurface, intra-tile control, tile gateway). Section III contains the conclusions and a brief discussion on the challenges and prospects of this emerging technological paradigm.

## II. THE HYPERSURFACE ARCHITECTURE AND FUNCTIONALITY

HSF tiles are envisioned as planar structures that can host metasurface functions with programmatic control. Each tile comprises a stack of virtual and physical components, shown in Fig. 1, the metasurface layer, the intra-tile control layer and the tile gateway layer. The HSF tile supports software descriptions of the metasurface functions and allows the programmer to customize, deploy or retract them on demand via a programming interface with appropriate callbacks with the general form: *outcome ← callback(action_type, parameters)*. The *action_type* is an identifier denoting the intended function to be applied to the impinging waves, e.g., STEER or ABSORB. Each action type is associated with a set of parameters. For instance, each function can be defined for specific: (i) incidence direction, (ii) steering direction, (iii) applicable frequency band, and (iv) polarization. The metasurface end-functionality is exposed to programmers via an API that serves as a strong layer of abstraction. It hides the internal complexity of the HSF and offers general purpose access to metasurface functions without requiring knowledge of the underlying hardware and physics. Thus, the reconfiguration matching the intended function is derived automatically without the programmer's intervention.

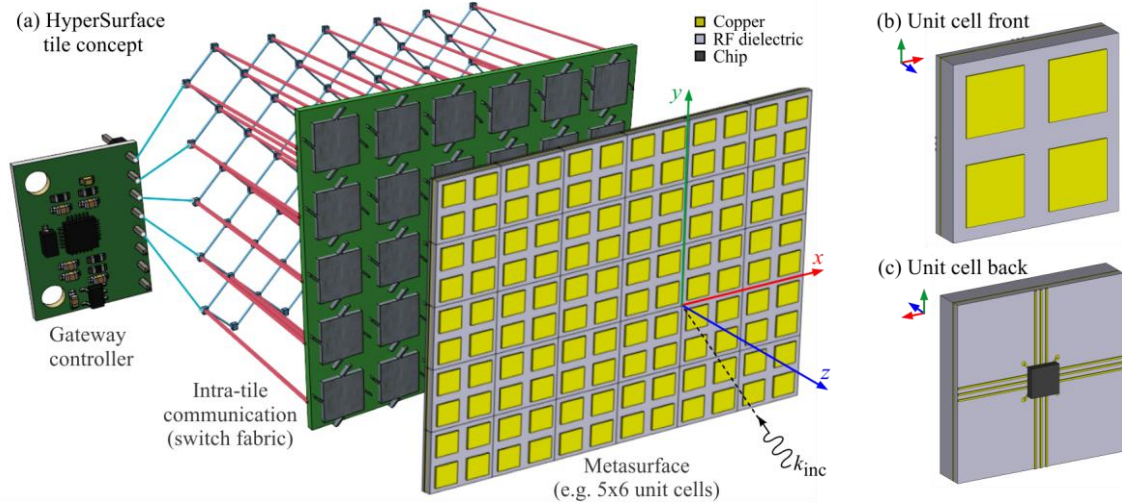

Fig. 1. (a) The functional and physical architecture of a single HyperSurface tile. The desired EM function is attained by a switch state configuration setup. The configuration commands are propagated within the tile via the controller intra-network. Inter-tile and external communication are handled by standard gateway hardware. Unit cell (b) front and (c) back depiction.

*The Metasurface Layer:* Two different EM implementations are examined for the HSF: (i) the switch-fabric approach based on a metallic-patch array and (ii) the graphene-based approach exploiting the tunable properties of graphene. The metasurface layer in the switch-fabric design consists of ultrathin copper patches periodically arranged in a dielectric, low-loss substrate (Rogers RT/Duroid 5880), back-plated by copper. The unit cell is small enough compared to the wavelength, in the order of a few mm, which ensures metasurface operation at the microwave regime, f=5 GHz. In the graphene-based approach we use periodically aligned square patches of Au placed on an Au-backed dielectric substrate. A graphene sheet is placed on the patches which effectively creates a patterning of the sheet conductivity. The size of the structure is in the μm scale and the operation frequency lies in the low THz spectrum, i.e., 2-15 THz, [3]. The response of the metasurface to the impinging plane wave is described by the complex-valued surface impedance, which consists of resistance (real part) and reactance (imaginary part) and it is a function of frequency and the geometrical and material parameters. The properties of the metasurface can be dynamically adjusted by making the complex-valued surface impedance tunable. The best option to achieve tunability is to use voltage-driven elements that control both the resistance and the reactance of the metasurface. To this end we implement a customized chip, the controller node, that hosts integrated circuitry capable of providing both variable resistance $R$ and capacitance $C$. Appropriate values, as required for each functionality, are propagated from the gateway via a network to the controllers, discussed below. Throughout our study we conduct analytical investigations and full-wave electromagnetic simulations utilizing established commercial software focusing on optimizing the designs with respect to the capabilities and restrictions of a realistic physical/circuit implementation of the controller nodes. By controlling the chip $RC$ values within the specified lumped electronic components, we demonstrate an angle-tunable absorber and advance towards more general scenarios of programmable wave control, including full-power reflection into any desired direction or directions (for instance, to create lenses with a tunable focus) [4,5] and environment-adaptive operation.

*The Intra-tile Control Layer*: The intra-tile control layer is composed of the electronic hardware component that enables the programmatic control over the HSF. It comprises a network of controllers, each with responsibility over one active meta-atom element. The control over a switching element is achieved via the control wiring shown in red, Fig. 1(a). Additionally, the controllers are able to exchange information in order to propagate switch configuration information within the tile. For example, a grid topology using separate wiring, e.g., shown as blue rhomb-pattern wires in Fig. 1(a), can be employed for the control tasks. The networked controller approach minimizes the required control wiring which benefits the manufacturing cost and the cross-talk minimization with the intended EM functionality. Furthermore, a fully wireless design for the interconnection of the controllers that employs energy harvesting can be envisaged.

*The Tile Gateway Layer*: The tile gateway layer specifies the hardware (Gateway) and protocols that enable the bidirectional communication between the controller network and the external world (such as the Internet) as well as the communication between the tiles. The interface and mapping between high level functionality and controller directives can also be handled by the Gateway. This provides flexibility and scalability in the HSF operation workflow. In general, multiple tiles are expected to be used as coating of large areas, operating as an EM "skin" that can be tuned locally and globally. Moreover, the tile hardware is intended to be inexpensive, favoring massive deployment. The networked tile architecture allows for offloading demanding tasks to powerful, external servers. A basic external service is responsible of deducing and returning the optimal tile configurations to be deployed after taking into consideration the status of other HSFs and their environment. Based on these specifications, existing IoT platforms can constitute exemplary choices for tile gateways [6].

## IV. CONCLUSION: CHALLENGES AND FUTURE PROSPECTS

The HyperSurface (HSF) is scalable planar metamaterial platform that bridges the gap between hardware and software, enabling programmatic control of the EM scattering properties over its surface. Meta-atoms are designed with specific metasurface functionalities in mind, which are tunable over a parameter range (e.g. operating frequency, incidence direction, polarization) and controllable by custom mm-sized chips constituting a nano-network. Each HSF tile is composed of a set of such tunable meta-atoms, e.g. $25 \times 25$, controlled by a single Gateway which serves as the interface between the software and hardware components of the platform, and handles the communication between other tiles and with the end-user (e.g. a computer). The HSF paradigm is a step towards the *Internet of Materials*, extending the metasurface concept, making it accessible to a wider audience and infusing it with novel functionality.

The ambition of the HSF paradigm comes with interesting challenges. From an EM design perspective, the tunability range of metasurfaces needs to be maximized following a co-design approach that takes into account the requirements of all the layers of the HSF. Another challenge the HSF design has to deal with is the intra- and inter-tile networking as the frequency increases. Various powering and communication schemes for the nodes (gateways or chips) are currently being researched [7], including network-on-chip architectures and wireless approaches, with the aim of quantifying the bitrate, latency/router delay, footprint, and consumption. Nevertheless the prospects of this new technology are manifold, ranging from adaptive "smart" metasurfaces & sensors to remotely-powered interactive EM mantles for IoT systems.

## ACKNOWLEDGEMENT

This work was supported by the European Union Horizon 2020 research and innovation programme-Future Emerging Topics (FETOPEN) under grant agreement No. 736876 (VISORSURF).